\documentclass[galaxies,article,accept,moreauthors,pdftex,10pt,a4paper]{mdpi}
\usepackage{footnote}

\usepackage{subfigure}
\makeatletter
\renewcommand{\@thesubfigure}{\normalsize(\textbf{\alph{subfigure}})}
\makeatother
\usepackage{soul}
\firstpage{1}
\articlenumber{x}
\doinum{10.3390/------}
\pubvolume{xx}
\pubyear{2016}
\externaleditor{Academic Editors:}
\history{}

 \theoremstyle{mdpi}
 \newcounter{thm}
 \newcounter{ex}
 \newcounter{re}
 \theoremstyle{mdpidefinition}

\Title{\textbf{Multiwavelength} Picture of the Blazar S5~0716+714 during Its Brightest Outburst}

\Author{Marina Manganaro $^{1,*}$, Giovanna Pedaletti $^{2}$, Marlene Doert $^{3}$, Denis Bastieri $^{4}$, \mbox{Vandad Fallah
Ramazani $^{5}$}, Dario Gasparrini $^{6}$, Elina Lindfors $^{5}$, Benoit Lott $^{7}$,
Mireia Nievas $^{8}$,  Bindu Rani $^{9}$, David J. Thompson $^{9}$, for the MAGIC and Fermi-LAT
Collaborations; \mbox{Emmanouil Angelakis $^{10}$}, George Borman $^{11}$,  Mark Gurwell $^{12}$, Talvikki Hovatta $^{13}$, \mbox{Ryosuke
Itoh $^{14}$}, Svetlana \textbf{Jorstad} $^{15}$, Alex Kraus $^{10}$, Thomas P. Krichbaum $^{10}$,
Paul Kuin $^{16}$, \mbox{Anne L\"{a}hteenm\"{a}ki $^{13,17}$}, Valeri Larionov $^{18}$,  Amy Yarleen Lien $^{9}$, Ioannis
Myserlis $^{10}$, \mbox{Merja Tornikoski $^{13}$}, Ivan Troitsky $^{18}$, J.  Anton Zensus $^{10}$}
\AuthorNames{Marina Manganaro, Giovanna Pedaletti, Marlene Doert, Denis Bastieri, Vandad Fallah
Ramazani, Dario Gasparrini, Elina Lindfors, Benoit Lott, Mireia Nievas,  Bindu Rani, David J. Thompson, for the MAGIC and \textit{Fermi}-LAT Collaborations; Emmanouil Angelakis, George Borman,  Mark Gurwell, Talvikki Hovatta, Ryosuke Itoh, Svetlana Jorstad,  Alex Kraus, Thomas P. Krichbaum, Paul Kuin, Anne L\"{a}hteenm\"{a}ki, Valeri Larionov,  Amy Yarleen Lien, Ioannis
Myserlis, Merja Tornikoski, Ivan Troitsky, J. Anton Zensus}

\address{%
$^{1}$ \quad Inst. de Astrof\'isica de Canarias, E-38200 La Laguna, and Universidad de La Laguna, Dpto. Astrof\'isica, E-38206 La Laguna, Tenerife, Spain\\
$^{2}$ \quad Deutsches Elektronen-Synchrotron (DESY), D-15738 Zeuthen , Germany; giovanna.pedaletti@desy.de   \\
$^{3}$ \quad Department of Physics, Technical University of Dortmund, Dortmund D-44221, Germany; marlene.doert@tu-dortmund.de\\
$^{4}$ \quad INFN Padova, I-35131 Padova, Italy; denis.bastieri@gmail.com\\
$^{5}$ \quad Department of Physics and Astronomy, University of Turku, 20500 Turku, Finland; vafara@utu.fi (V.F.R.); elilin@utu.fi (E.L.)\\
$^{6}$ \quad ASI Science Data Center and INFN, 06123 Perugia, Italy; dario.gasparrini@asdc.asi.it\\
$^{7}$ \quad CEN Bordeaux-Gradignan, 33170 Gradignan, France; lott@cenbg.in2p3.fr\\
$^{8}$ \quad Department of Atomic Molecular and Nuclear Physics, Universidad Complutense, E-28040 Madrid , Spain; mnievas@ucm.es\\
$^{9}$ \quad NASA's Goddard Space Flight Center, MD 20771 Greenbelt, USA; bindu.rani@nasa.gov (B.R.); David.J.Thompson@nasa.gov (D.J.T.); amy.y.lien@nasa.gov (A.Y.L.)\\
$^{10}$ \,\, MPIfR Max Planck Institut f\"{u}r Radioastronomie, 53121 Bonn, Germany; eangelakis@mpifr.de (E.A.); akraus@mpifr.de (A.K.); tkrichbaum@mpifr-bonn.mpg.de (T.P.K.); imyserlis@mpifr-bonn.mpg.de (I.M.); azensus@mpifr-bonn.mpg.de (J.A.Z.)\\
$^{11}$ \,\, Crimean Astrophysics Observatory, 98409 Nauchny, Crimea, Russia; borman.ga@gmail.com\\
$^{12}$ \,\, Harvard-Smithsonian CfA, MA 02138 Cambridge, USA; mgurwell@cfa.harvard.edu\\
$^{13}$ \,\, Mets\"{a}hovi Radio Observatory, Aalto University, FI-02540 Kylm\"{a}l\"{a}; talvikki.hovatta@aalto.fi (T.H.);  \protect\linebreak anne.lahteenmaki@aalto.fi (A.L.); merja.tornikoski@aalto.fi (M.T.)\\
$^{14}$ \,\, Department of Physical Science, Hiroshima University, Higashi-hiroshima, 739-8526, Japan; itoh@hep01.hepl.hiroshima-u.ac.jp\\
$^{15}$ \,\, Institute for Astrophysical Research, Boston University, Boston MA 02215; jorstad@bu.edu\\
$^{16}$ \,\, Mullard Space Science Lab., UCL, RH5 6NT Dorking, UK; n.kuin@ucl.ac.uk\\
$^{17}$ \,\, Department of Radio Science and Engineering, Aalto University, FI-00076 Aalto, Finland;\\
$^{18}$ \,\, Astronomical Institute, St.-Petersburg State University, St.-Petersburg 198504, Russia; \mbox{v.larionov@spbu.ru (V.L.)}; i.troitsky@spbu.ru (I.T.)
}
\corres{Correspondence: manganaro@iac.es; Tel.: +34-922-605-551}

\abstract{S5~0716+714 is a well known {BL 
		 Lac} object, and one of the brightest and most active blazars. The discovery in the Very High Energy band (VHE, E > 100 GeV) by MAGIC happened in 2008. In~January 2015, the source went through the brightest optical state ever observed, triggering MAGIC follow-up and a VHE detection with $\sim13\sigma$ significance (ATel $\sharp6999$). Rich multiwavelength coverage of the flare allowed us to construct the broad-band spectral energy distribution of S5~0716+714 during its brightest outburst. In this work, we will present the preliminary analysis of MAGIC and \textit{Fermi}-LAT 
	  data of the flaring activity in January and February 2015 for the HE (0.1 < HE < 300 GeV) and VHE band, together with radio (Mets\"{a}hovi, OVRO, VLBA, Effelsberg), sub-millimeter (SMA), optical (Tuorla, Perkins, Steward, AZT-8+ST7, LX-200, Kanata), X-ray and UV 
	  (\textit{Swift}-XRT and UVOT), in the same time-window and discuss the time variability of the {multiwavelength} light curves during this impressive outburst.\\[12pt]}


\conferencetitle{"Blazars through sharp MW eyes"}

\begin{document}

\section{Introduction}
The BL Lac object S5~0716+714 is among the  brightest and most active blazars \cite{Rani2013}. Because  of  the  featureless  optical  continuum,  it  is  hard  to  estimate  its  redshift. A value of $z= 0.31 \pm0.08$ is derived from the photometric detection  of  the  host  galaxy \cite{nilsson2008}. It was detected in VHE  for the first time by MAGIC in 2008 \cite{Anderhub2009}. Because of its extreme variability, S5~0716+714 has been the subject of several optical monitoring campaigns (e.g., \cite{wagner1996, rani2011, montagni2006, Rani2013, rani2015} and references therein). S5~0716+714 is also among the bright blazars in the \textit{Fermi}-LAT (Large Area Telescope) Bright AGN  Sample (LBAS) \cite{abdo2010LBAS}, whose GeV spectra are governed by a broken power law. The combined GeV--TeV spectrum of the source displays absorption-like features in 10--100~GeV energy range {\cite{senturk2013}}. The blazar S5~0716+714 was observed in an unprecedented outburst phase during January 2015. The source was detected at its historic high brightness at optical and IR  bands. In January 2015, MAGIC, triggered by the high optical state and by high energy photons detected by \textit{Fermi}-LAT, detected the source with a significance of $\sim 13\sigma$ \cite{Atel6999}.

\section{Instruments Involved}
MAGIC is a stereoscopic system consisting of two 17\,m diameter Imaging Atmospheric Cherenkov Telescopes located at the Observatorio del Roque de los Muchachos, on the Canary Island of La Palma (Spain). 
The current integral sensitivity for the zenith range $35^{\circ}<zd<50{^\circ}$ above 100\,GeV is $1.76\%\pm0.03\%$ of the Crab Nebula's flux in 50~h. The Field of View (FoV) of MAGIC is $3.5^{\circ}$ \cite{Aleksic2015_Upgrade2}.
The Large Area Telescope (LAT) on board of {\it Fermi} is an imaging high-energy $\gamma$-ray telescope in the energy range from about 20 MeV to more than 300 GeV \cite{Atwood2009}. The LAT's Field of View covers about 20\% of the sky at any time, and it scans continuously, covering the whole sky every three hours.
Multiwavelength observations were performed  collecting data from a large number of instruments.
Our dataset includes data from \textit{Swift}-XRT \cite{Burrows2005} in the X-ray and \textit{Swift}-UVOT \cite{Roming2005} in optical-UV; for the optical band we show data from the Tuorla Blazar monitoring program {(http://users.utu.fi/kani/1m)}, the  1.8 meter Perkins telescope {(https://lowell.edu/research/research-facilities/1-8-meter-perkins/)}, the Steward Observatory {(http://www.bu.edu/blazars/)}, the 70-cm AZT-8 reflector of the Crimean Astrophysical Observatory {(http://craocrimea.ru/ru)}, the 40-cm LX-200 telescope in St. Petersburg, Russia, \cite{HagenTorn2006} and  Kanata telescope at the Hiroshima Observatory {(http://hasc.hiroshima-u.ac.jp/telescope/kanatatel-e.html)}; radio band was covered by Mets\"{a}hovi~\cite{Terasranta}, the Owens Valley Radio Observatory (OVRO) 40~m telescope \cite{Richards2011} and the Effelsberg 100~m radio telescope {(http://www.mpifr-bonn.mpg.de/en/effelsberg)}. Radio data at 230 GHz band are obtained using the Submillimeter Array {(http://sma1.sma.hawaii.edu/)} (SMA).

\section{Observations and Results}
\vspace{-6pt}

\subsection{MAGIC Analysis}

\begin{figure}[H]
\centering
\subfigure[~Distribution of the squared angular distance $\theta^{2}$]{%
\includegraphics[width=17pc]{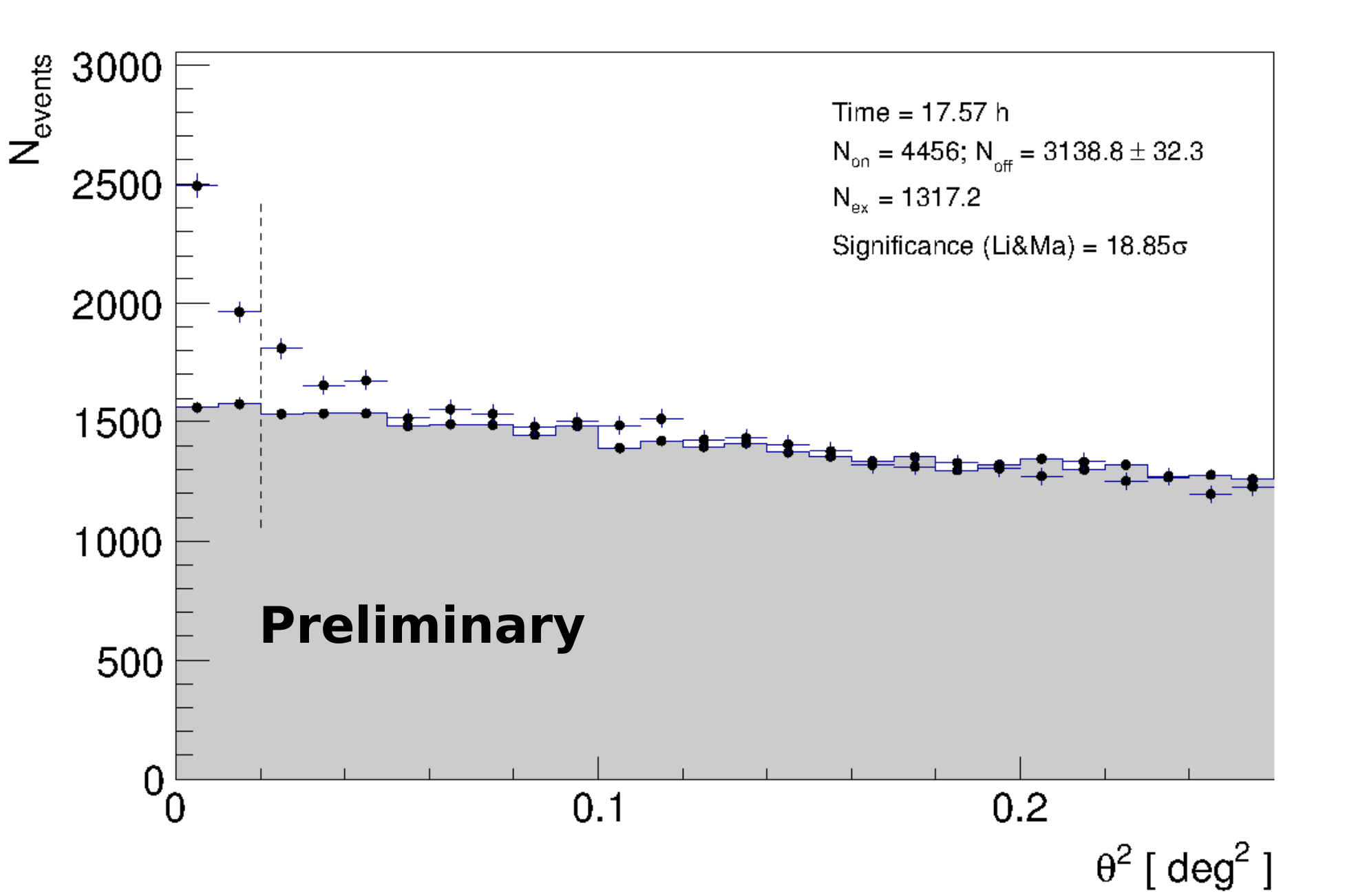}
\label{fig:th2}}
\quad
\subfigure[~Skymap]{%
\includegraphics[width=14pc]{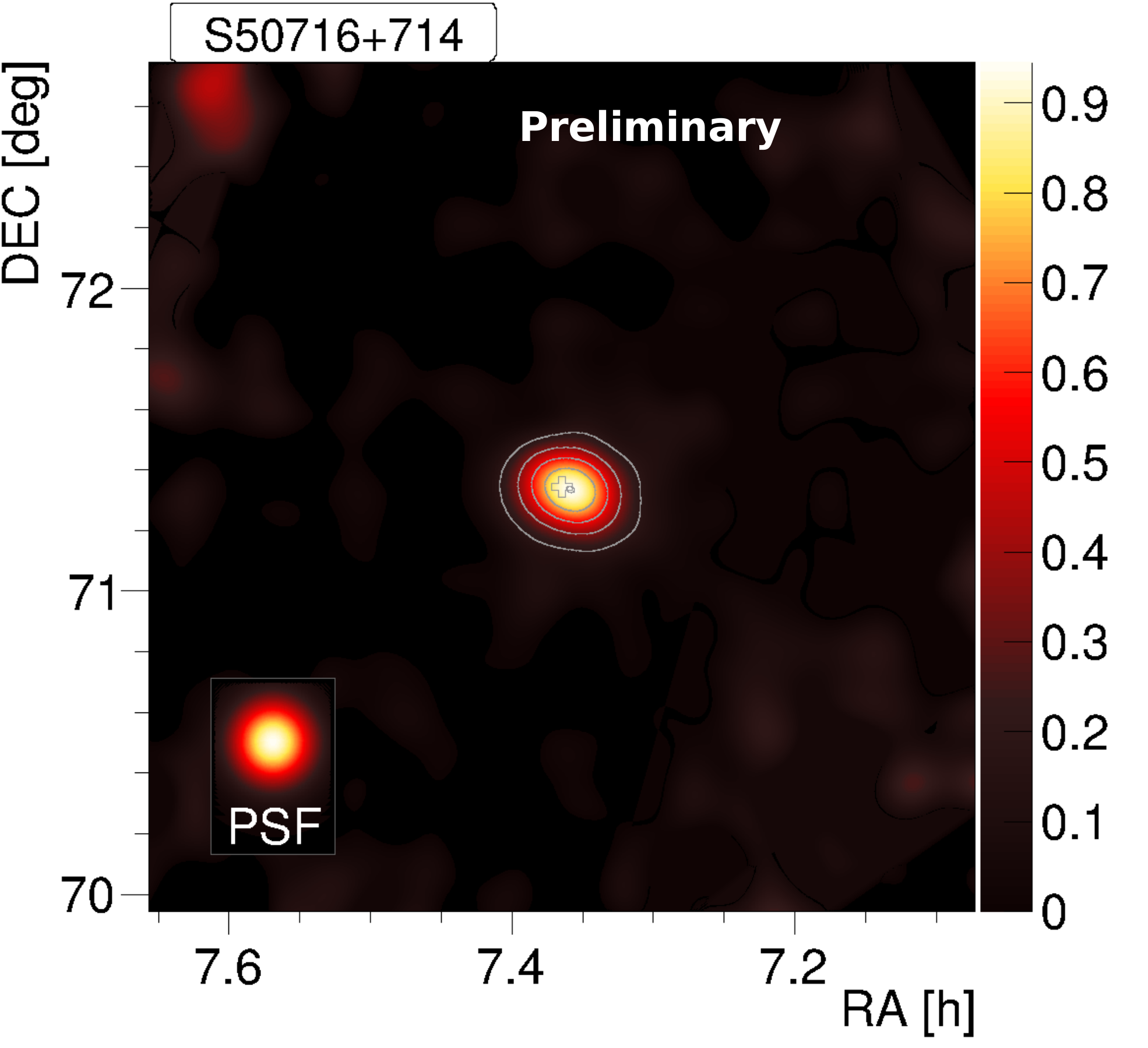}
\label{fig:skymap}}
\caption{Distribution of the squared angular distance $\theta^{2}$ between the reconstructed source position and the nominal source position (points) or the background estimation position (shaded area) for S5~0716+714 (\textbf{a}) and VHE skymap of the source (\textbf{b}). The color bar indicates the relative flux. The~contours indicate significance in steps of $5 \sigma$.}
\label{fig:MAGIC}

\end{figure}

MAGIC started to observe S5~0716+714 on 19 January {2015}. Unfortunately, the following two nights of strong wind during the S5~0716+714 visibility window prevented {observation} of the source, which was resumed on 23 January. From that day, the flux constantly increased up to the maximum flux ever observed in the VHE range for this source, on 27 January. \linebreak The next activity of S5~0716+714 in the VHE was detected by MAGIC on the 13 February, and this time lasting three days only, up to the 16 February. {In the present work, we will refer to January data (from MJD 	 57040 to MJD 57050) as phase A and to February data (from MJD 57066 to MJD 57069) as phase B}. The observations were performed in wobble mode with a $0.4\deg$ offset and four symmetric positions with respect to the camera center. The observations were performed in 12 nights in January and February 2015 for a total duration of {17.57~h} at the zenith range of $40^\circ<zd<50^\circ$). The corresponding distribution of the squared angular distance $\theta^{2}$ between the reconstructed source position and the nominal source position (points) or the background estimation position (shaded area) for the whole period of observation is shown in Figure~\ref{fig:MAGIC}a. The vertical dashed line shows the value of $\theta^{2}$ up to which the number of excess events and significance are integrated. The cuts used are optimized for steep {spectra} and low energy thresholds. The analysis energy threshold is \mbox{$\sim$ 65 GeV}, measured as the peak of the energy distribution of a sample of Monte Carlo simulated events for a source with {the} spectral shape of S5~0716+714. The analysis of the data collected was performed within the standard MAGIC analysis framework \emph{MARS} \cite{Zanin2013_MARS,Aleksic2015_Upgrade2}.

\subsection{\textit{Fermi}-LAT Analysis}

The HE $\gamma$-ray (0.1--300~GeV) observations were obtained in a survey mode by the {\it Fermi}-LAT Large Area Telescope. The LAT data  were analyzed using the standard ScienceTools (software version v10.01.01) and instrument response function CALDB. Photons in the source event class were selected for the analysis. We analyzed a region of interest of 10$^{\circ}$ in radius centred at the position of S5 0716+714 using a maximum-likelihood algorithm \cite{mattox1996}. We included all the 54 sources of the 3FGL catalogue~\cite{Acero2015} within 10$^{\circ}$ in the unbinned likelihood analysis. Model parameters for sources within 5 degrees of the ROI 
are kept {free, while} we kept the model parameters for the rest fixed to their catalogue values.

\subsection{{Multiwavelength} Light Curves}

\begin{figure}[H]
\centering
\includegraphics[width=13cm]{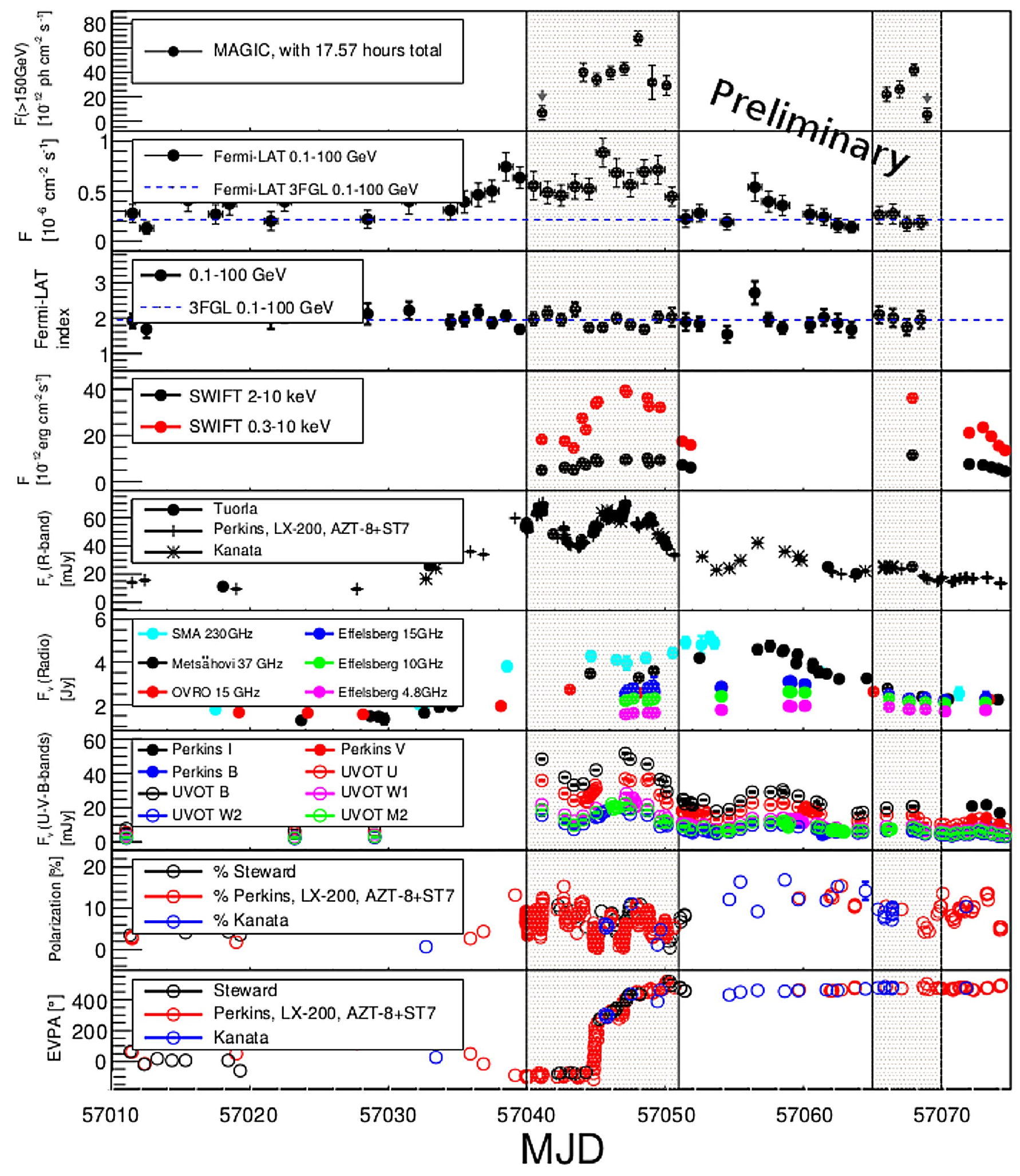}
\caption{{Multiwavelength light curves of S5 0716+714. Shadowed areas indicate Phase A (from MJD 57040 to MJD 57050) and Phase B (from MJD 57066 to MJD 57069) and the respective activities in all the energy bands. \textit{Fermi}-LAT panels (second and third from the top) show the 3FGL averaged flux and spectral index from the 3FGL Catalog \cite{Acero2015}, respectively, as blue dashed lines.}}
\label{fig:MWL}
\end{figure}

The unprecedented outburst of S5~0716+714 was covered in all the wavelengths: such a rich collection of simultaneous data allows us to study with great precision the broadband spectral energy distribution of the source, deeply investigating its nature. {Multiwavelength} flux light curves of S5~0716+714 during the period from MJD 57010 to MJD 57075 are shown in Figure~\ref{fig:MWL}. The top panel shows the MAGIC light curves: in the VHE band, the no-variability hypothesis has been discarded because it resulted in a $\chi^2/ndf=67/7$ for January and  $\chi^2/ndf=22/3$ for the February data for a constant fit. Intranight variability in VHE was not detected. The shadowed areas indicate Phase A (from MJD 57040 to MJD 57050) and Phase B (from MJD 57066 to MJD 57069), the two high states in VHE, and the corresponding activity in the other bands.

When MAGIC discovered S5~0716+714 in 2008 (\cite{Anderhub2009}) the radio band was in a quiescent level, while here, even if delayed with respect to the optical peak, it is possible to discern a response in the radio frequency. {The VHE integral flux measured in 2008 by MAGIC was 10 and six times lower with respect to the one detected during the phase A and phase B outbursts, respectively.}
The two phases show quite different behavior: phase A  is characterized by a high state of the source clearly visible in every energy band, especially in optical where the highest flux ever detected for S5~0716+714 is reported. {HE and} optical curves show a double peaked structure and the polarization angle undergoes a fast change between the two optical peaks.
After an intermediate time period when activity is limited to the radio band, phase B is characterized by VHE and X-ray high flux, while the other bands reached a lower state after the outburst of phase A. The fractional variability of both phases shows an increasing trend with the energy. The phase B flare is clearly visible in the VHE and X-ray band only. Such differences in behaviour between the two activity states is very interesting to study, and together with the jet analysis of the source and the modeling of the broadband spectral energy distribution, which will be presented in \cite{preparation}, it will improve our understanding  of the {BL Lac} emission mechanism~\cite{Rani2016}.

\section{Discussion}
The unprecedented outburst of the {BL Lac} object S5~0716+714 in January 2015 was observed in all the wavebands. The MAGIC detection of the source, together with simultaneous data in the HE range from \textit{Fermi}-LAT, will allow us to constrain the Inverse Compton part of the broadband spectrum. The rich multiwavelength coverage presented here is part of a more comprehensive study of the source in preparation \cite{preparation}. The impressive activity in phase A was followed by some days of decreasing flux, and then a new high-state was observed in the VHE and X-ray bands (phase B). The structure of the light curve in phase A showed a double-peaked shape, and the electric vector position angle (EVPA) of the linearly polarized optical emission  rotated by $~\sim 360^{\circ}$ in between the two~peaks.
Such swings in EVPA have been frequently observed in connection with $\gamma$-ray flaring epochs \linebreak(e.g., \cite{Marscher2010,Aleksic2014,Blinov2015}). A detailed correlation study and modelling of the spectral energy distribution are in~progress.

\vspace{6pt}


\acknowledgments{We would like to thank the Instituto de Astrof\'{\i}sica de Canarias
for the excellent working conditions at the Observatorio del Roque de los Muchachos in La Palma. The financial support of the German BMBF and MPG, the Italian INFN and INAF, the Swiss National Fund SNF,
the he ERDF under the Spanish MINECO (FPA2015-69818-P, FPA2012-36668, FPA2015-68278-P,FPA2015-69210-C6-2-R, FPA2015-69210-C6-4-R,
FPA2015-69210-C6-6-R, AYA2013-47447-C3-1-P, AYA2015-71042-P, ESP2015-71662-C2-2-P, CSD2009-00064), and the Japanese JSPS and MEXT
is gratefully acknowledged. This work was also supported by the Spanish Centro de Excelencia ``Severo Ochoa'' SEV-2012-0234 and SEV-2015-0548, and Unidad de Excelencia ``Mar\'{\i}a de Maeztu'' MDM-2014-0369, by grant 268740 of the Academy of Finland,
by the Croatian Science Foundation (HrZZ) Project 09/176 and the University of Rijeka Project 13.12.1.3.02, by the DFG Collaborative Research Centers SFB823/C4 and SFB876/C3, and by the Polish MNiSzW grant 745/N-HESS-MAGIC/2010/0.
The \textit{Fermi}-LAT Collaboration acknowledges support for LAT development, operation and data analysis from NASA and DOE (United States), CEA/Irfu and IN2P3/CNRS (France), ASI and INFN (Italy), MEXT, KEK, and JAXA (Japan), and the K.A. Wallenberg Foundation, the Swedish Research Council and the National Space Board (Sweden). Science analysis support in the operations phase from INAF (Italy) and CNES (France) is also gratefully acknowledged. We thank the {\it Swift} team duty scientists and science planners. The Mets\"ahovi team acknowledges the support from the Academy of Finland to our observing projects (numbers 212656, 210338, 121148, and others). The Submillimeter Array is a joint project between the Smithsonian Astrophysical Observatory and the Academia Sinica Institute of Astronomy and Astrophysics and is funded by the Smithsonian Institution and the Academia Sinica. The OVRO 40-m monitoring program is supported in part by NASA grants NNX08AW31G, NNX11A043G and NNX14AQ89G, and NSF grants AST-0808050 and AST-1109911. St.-Petersburg University team  was supported by research grant 6.38.335.2015 and RFBR grant 15-02-00949.}

\authorcontributions{Marina Manganaro, Giovanna Pedaletti and Mireia Nievas performed the MAGIC analysis of the source. Bindu Rani, Denis Bastieri, Dario Gasparrini took care of the \textit{Fermi}-LAT analysis. This proceeding is part of a joint MAGIC-\textit{Fermi}-LAT work involving also other multiwavelength authors. The preliminary results here included will be part of a paper in preparation which is led by Marina Manganaro, Marlene Doert and Bindu Rani. All the other authors listed provided the analysis of their data and participated in the discussion of the {multiwavelength} curve. Elina Lindfors is the PI of the MAGIC program of ToO observations of flaring AGNs based on optical, X-ray and $\gamma$-ray triggers. Marina Manganaro wrote this proceeding.}

\conflictofinterests{The authors declare no conflict of interest.}

\bibliographystyle{mdpi}

\begin{thebibliography}{------}

\bibitem{Rani2013}
Rani, B.;   Krichbaum, T.P.;  Fuhrmann, L.;    Boettcher, M.;   Lott, B.;   Aller, H.D.;  Aller, M.F.;   Angelakis, E.;  Bach,~U.;  Bastieri, D.; et al.  Radio to gamma-ray variability study of blazar S5 0716+714. {\em Astron. Astrophys.} {\bf 2013}, {\em 552}, A11.

\bibitem{nilsson2008}
Nilsson, K.;  Pursimo, T.;   Sillanpää, A.;  Takalo, L.O.;  Lindfors, E. Detection of the host galaxy of S5~0716+714. {\em Astron. Astrophys.} {\bf 2008}, {\em 487}, L29--L32.


\bibitem{Anderhub2009}
Anderhub, H. Discovery of very high energy $\gamma$-rays from the blazar S5~0716+714. {\em Physics} {\bf 2009}, {\em 704}, L129--L133.


\bibitem{wagner1996}
Wagner, S.J.; Witzel, A.; Heidt, J.; Krichbaum, T.P.; Qian, S.J.; Quirrenbach, A.; Wegner, R.; Aller, H.; Aller,~M.; Anton, K.; et al. Rapid variability in S5~0716+714 across the electromagnetic spectrum.  {\em Astron. J.}  {\bf 1996}, {\em 111}, 2187--2211.

\bibitem{rani2011}
Rani, B.; Gupta, A.C.; Joshi, U.C.; Ganesh, S.; Wiita, P.J. Optical intraday variability studies of 10 low energy peaked blazars. {\em Mon. Not. R. Astron. Soc.} {\bf 2011}, {\em 413}, 2157--2172.

\bibitem{montagni2006}
Montagni, F.;  Maselli, A.;   Massaro, E.;   Nesci, R.;   Sclavi, S.;   Maesano, M. The intra-night optical variability of the bright BL Lacertae object S5~0716+714. {\em Astron. Astrophys.} {\bf 2006}, {\em 451}, 435--442.


\bibitem{rani2015}
Rani, B.;  Krichbaum, T.P.;  Marscher, A.P.;  Hodgson, J.A.;  Fuhrmann, L.;  Angelakis, E.;  Britzen, S.;  Zensus,~J.A.  Connection between inner jet kinematics and broadband flux variability in the BL Lacertae object S5~0716+714. {\em Astron. Astrophys.} {\bf 2015}, {\em 578}, A123.



\bibitem{abdo2010LBAS}
Abdo, A.A.; Ackermann, M.;   Ajello, M.;   Atwood4, W.B.;  Axelsson, M.;   Baldini, L.;   Ballet, J.;   Barbiellini, G.;   Bastieri, D.;   Bechtol, K.; et al.  Spectral Properties of bright Fermi-detected blazars in the gamma-ray band. {\em Astrophys. J.} {\bf 2010}, {\em 710}, 1271.

\bibitem{senturk2013}
Senturk, G.D.;  Errando, M.; Böttcher, M.; Mukherjee, R. Gamma-ray Observational properties of TeV-detected blazars. {\em Astrophys. J.}  {\bf 2013}, {\em 764}, 119.



\bibitem{Atel6999}
Mirzoyan, R. MAGIC detects Very High Energy gamma-rays from S5 0716+714. Available online: http://www.astronomerstelegram.org/?read=6999 (accessed on 21 November 2016).


\bibitem{Aleksic2015_Upgrade2}
Aleksi\'c, J.;  Ansoldib, S.;  Antonellic, L.A.;  Antoranzd, P.;  Babice, A.;  Bangalef, P.;  Barcelóa, M.;  Barriog, J.A.;  Gonzálezh, J.B.;   Bednarek, W.; et al. The major upgrade of the MAGIC telescopes, Part II: A performance study using
observations of the Crab Nebula. {\em Astropart. Phys.} {\bf 2016}, {\em 72}, 76--94.

\bibitem{Atwood2009}
Atwood, W.B.;  Abdo, A.A.;   Ackermann, M.;   Althouse, W.;   Anderson, B.;   Axelsson, M.;   Baldini, L.;   Ballet,~J.;   Band, D.L.;    Barbiellini, G.; et al. The Large Area Telescope on the Fermi Gamma-Ray Space Telescope Mission. {\em Astrophys. J.} {\bf 2009}, {\em 697}, 1071.

\bibitem{Burrows2005}
Hill, J.E.; Burrows, D.N.; Nousek, J.A.; Wells, A.; Turner, M.; Willingale, R.; Holland, A.; Citterio, O.; Chincarini, G.; Campana, S.; et al. The \textit{Swift} X-ray telescope. {\em Space Sci. Rev.} {\bf 2005}, {\em 120}, 165--195.


\bibitem{Roming2005}
Roming, P.W.A.; Hunsberger, S.D.; Nousek, J.A.; Mason, K.O.; Breeveld, A.A. The \textit{Swift} Ultra-Violet/Optical Telescope. {\em Space Sci. Rev.} {\bf 2005}, {\em 120}, 95--142.


\bibitem{HagenTorn2006}
Hagen-Thorn, V.A.;  Larionov, V.M.;  Efimova, N.V.;  Hagen-Thorn, E.I.;  Arkharov, A.A.;  Di Paola, A.;  Dolci,~M.;  Takalo, L.O.;  Sillanpää, A.;  Ostorero, L.; et al. Optical and IR monitoring of the BL Lac object S5 0716+714 from 2001--2004. {\em Astr. Rep.} {\bf 2006}, {\em 50}, 458.



\bibitem{Terasranta}
Ter\"asranta, H.;  Tornikoski, M.; Mujunen, A.; Karlamaa, K.; Valtonen, T.; Henelius, N.; Urpo, S.; Lainela, M.; Pursimo, T.; Nilsson, K.; et al. Fifteen years monitoring of extragalactic radio sources at 22, 37 and 87 GHz. {\em Astron. Astrophys. Suppl.} {\bf 1998}, {\em 132}, 305--331.

\bibitem{Richards2011}
Richards, J.L.;   Max-Moerbeck, W.;   Pavlidou, V.;   King, O.G.;   Pearson, T.J.;   Readhead, A.C.S.;  Reeves, R.;   Shepherd, M.C.;   Stevenson, M.A.;   Weintraub, L.C.; et al. Blazars in the Fermi era: the OVRO 40 m telescope monitoring program. {\em Astrophys. J. Suppl.} {\bf 2011}, {\em 194}, 29.

\bibitem{Zanin2013_MARS}
Zanin, R.;  Gaug, M.;  Carmona, E.;  Colin, P.;  Delgado, C.;  Lombardi, S.;  Mazin, D.;  Scalzotto, V.;  Sitarek,~J.;  Tescaro, D.; et al.  MARS, the MAGIC analysis and reconstruction software. {In Proceedings of the 33th International Cosmic Ray Conference (ICRC)}, Rio de Janeiro, Brazil, 2--9 July 2013;  {Volume 773}.

\bibitem{mattox1996}
Mattox, J.R.; Bertsch, D.L.; Chiang, J.; Dingus, B.L.; Digel, S.W.; Esposito, J.A.; Fierro, J.M.; Hartman, R.C.; Hunter, S.D.; Kanbach, G.; et al. The likelyhood analysis of EGRET data. {\em Astrophys. J.} {\bf 1996},  {\em 461}, 396.

\bibitem{Acero2015} 	
Acero, F.; Ackermann, M.; Ajello, M.; Albert, A.; Atwood, W. B.; Axelsson, M.; Baldini, L.; Ballet, J.; Barbiellini, G.; Bastieri, D.; et al.\textit{Fermi} Large Area Telescope Third Source Catalog. {\em Astrophys. J. Suppl.} {\bf 2015}, {\em 218}, 23.



\bibitem{preparation}
Ahnen, M.L.; Ahnen, M.~L., Ansoldi, S., Antonelli, L.~A., Arcaro, C., Babi\'c, A., Banerjee, B., Bangale, P., Barres de Almeida, U., Barrio, J.~A.,Becerra Gonz\'alez, J.; {et al.} MWL characterization of the blazar S5~0716+714  during an unprecedented outburst phase. 2016, in preparation.


\bibitem{Rani2016}

Rani, B.;   Krichbaum, T.P.;  Hodgson, J.A.;  Zensus, J.A. Location and origin of gamma-rays in blazars.\linebreak {\em J. Phys. Conf. Ser.},  {\bf 2016}, {\em 718}, 052032.


\bibitem{Marscher2010}
Marscher, A.;   Marscher, A.P.;   Jorstad, S.G.;   Larionov, V.M.;   Aller, M.F.;   Aller, H.D.;  Lähteenmäki, A.;  Agudo,~I.;   Smith, P.S.;  Gurwell, M.; et al. Probing the inner jet of the quasar PKS 1510-089 with multi-waveband monitoring during strong gamma-ray activity. {\em  Astrophys. J. Lett.} {\bf 2010}, {\em 710}, L126--131.

\bibitem{Aleksic2014}
Aleksi\'c, J.;   Ansoldi, S.;   Antonelli, L.A.;   Antoranz, P.;    Babic, A.;    Bangale, P.;   Barres de Almeida,~U.;   Barrio,~J.A.;   Becerra González, J.;   Bednarek, W.;   Bernardini, E.; et al. MAGIC gamma-ray and multi-frequency observations of flat spectrum radio quasar PKS 1510-089 in early 2012. {\em Astron. Astrophys.}  {\bf 2014}, {\em 569},~A46.


\bibitem{Blinov2015}
Blinov, D.;   Pavlidou, V.;   Papadakis, I.;   Kiehlmann, S.;   Panopoulou, G.;   Liodakis, I.;   King, O.G.;   Angelakis,~E.;   Baloković, M.;   Das, H.; et al. RoboPol: First season rotations of optical polarization plane in blazars. {\em  Mon. Not. R. Astron. Soc.}  {\bf 2015}, {\em 453}, 1669--1683.



\end{thebibliography}

\renewcommand\bibname{References}

\end{document}